\documentclass{article}
\usepackage{icmc2017template}
\usepackage{times}
\usepackage{ifpdf}
\usepackage{soul}
\usepackage[english]{babel}
\usepackage{amsmath}
\usepackage{url}

\usepackage{enumitem}
\setlist{noitemsep}  

\def\papertitle{Composition by Conversation}
\def\firstauthor{Donya Quick}
\def\secondauthor{Clayton T. Morrison}
\def\thirdauthor{Third Author}

\newif\ifpdf
\ifx\pdfoutput\relax
\else
   \ifcase\pdfoutput
      \pdffalse
   \else
      \pdftrue
  \fi
\fi

\ifpdf 
  \usepackage[pdftex,
    pdftitle={\papertitle},
    pdfauthor={\firstauthor, \secondauthor, \thirdauthor},
    bookmarksnumbered, 
    pdfstartview=XYZ 
   ]{hyperref}

  \usepackage[pdftex]{graphicx}
  \graphicspath{{./figures/}}
  \DeclareGraphicsExtensions{.pdf,.jpeg,.png}

  \usepackage[figure,table]{hypcap}

\else 
  \usepackage[dvips,
    bookmarksnumbered, 
    pdfstartview=XYZ 
  ]{hyperref}  

  \usepackage[dvips]{epsfig,graphicx}
  \graphicspath{{./figures/}}
  \DeclareGraphicsExtensions{.png}

  \usepackage[figure,table]{hypcap}
\fi

\hypersetup{
    colorlinks,%
    citecolor=black,%
    filecolor=black,%
    linkcolor=black,%
    urlcolor=black
}

\usepackage{xspace}
\long\def\museci{MusECI\xspace}

\title{\papertitle}

 \twoauthors
   {1.5in}
   {\firstauthor} {Stevens Institute of Technology \\  
     {\tt \href{mailto:dquick@stevens.edu}{dquick@stevens.edu}}}
   {\secondauthor} {University of Arizona \\  
    {\tt \href{mailto:claytonm@email.arizona.edu}{claytonm@email.arizona.edu}}}

\begin{document}
\capstartfalse
\maketitle
\capstarttrue
\begin{abstract}
Most musical programming languages are developed \linebreak purely for coding virtual instruments or algorithmic compositions. Although there has been some work in the domain of musical query languages for music information retrieval, there has been little attempt to unify the principles of musical programming and query languages with cognitive and natural language processing models that would facilitate the activity of composition by conversation. We present a prototype framework, called \museci, that merges these domains, permitting score-level algorithmic composition in a text editor while also supporting connectivity to existing natural language processing frameworks.
\end{abstract}

\section{Introduction}


We have reached an age where natural language interfaces are increasingly available for our technology. These interfaces provide opportunities to interact more naturally with our devices. For example, it is becoming common for people to speak directly to mobile devices to accomplish tasks with no human on the other end of the interactions. However, to date there has been little work on the support of conversation and interactions involving musical concepts and scores.

Deep learning with neural networks is currently a popular strategy in the domain of natural language processing (NLP), and has been applied to developing dialog systems where a machine attempts to carry on a conversation with a human \cite{nn-dialog-systems}. 
However, neural nets remain difficult to analyze and extend once trained.
In contrast, the classic approach of explicit \emph{knowledge representation} symbolically represents ideas and thought processes that can be directly analyzed and extended, and provides a natural framework for explaining decisions.


Falling under the general strategy of knowledge representation, elementary composable ideas, or ECIs, are simple, atomic concepts that can be combined, or \emph{composed}, to form more complex concepts and relationships. For example, there is some evidence to suggest that ``surface'' is one such concept that infants have and use to build other concepts, such as collections of surfaces forming objects and internal spaces \cite{mandler2004}. This style of representation has been proposed to underly human natural language processing tasks such as describing and referring to items and events in visual environments \cite{stamant2006,voxml}. 

Programming languages already follow an ECI-like \linebreak model: there are primitives, such as integer and floating-point numbers, and a collection of basic operations and data structures over those entities that are used to represent more complex concepts. Object-oriented programming seeks to model computational problems much in the way we view our environment as composed of objects, each of which consists of parts that interact and can be manipulated in particular ways. However, not all ideas fit neatly into this approach and there are many languages that use very different strategies, such as functional programming languages. Additionally, appropriate language features for a computer are not necessarily straightforward derivations from verbal descriptions---there is a reason why programming a computer is a complex skill that requires specialized training.

We propose a basic collection of ECIs for music with an accompanying Python implementation called \museci
\footnote{Our implementation is available at \url{musica.ml4ai.org/}.}. 
In this paper we describe ECI-style musical primitives and structures for grouping these primitives. We also provide basic operations to facilitate conversational manipulation of music through a query-then-operation pattern of interaction. 
This lays the foundation for eventually building larger platforms that would facilitate real-time conversations with machines about music. The goal of \museci is not to replace traditional score editing software, but rather to enable the creation of new systems to augment those environments and open new avenues for human-computer interaction through music.

\section{Programming Languages}

A central goal of the \museci project is to create a language for score-level music representation that can be used in an algorithmic or automated composition setting while still adhering closely to models of music cognition and expression in natural language. 
We adopt the constraints that the data structures used to represent music both have a straightforward interpretation for editing and playback and that they can also be easily queried to identify features based on natural language descriptions and manipulated according to instructions.


Many musical programming languages focus on the \linebreak sound, or signal-level aspects of music generation. Examples of these include Csound \cite{csound}, SuperCollider\cite{supercollider}\footnote{SuperCollider also represents another language paradigm known as live coding\cite{live-coding}, which allows compilation and execution of fragments of a larger program on demand. However, while gaining traction in music performance, this kind of language paradigm has not been interfaced with NLP systems.}, 
and Max/MSP\cite{cycling74}. While these languages are extremely useful in the context of music creation, they do not offer a rich variety of representations for higher level features in music. On the other extreme are languages that focus on creating visually appealing musical scores.  Languages such as LilyPond \cite{lilypond} focus on type-setting aspects of a score, with emphasis on details of visual presentation and formatting and less on relationships between musical features that would be useful for modeling music cognition. We seek a representation between the extremes of visual detail and sound synthesis, focusing on common natural language descriptors people use to talk about music: note, melody, chord, phrase, and so on. 

Other languages for representing score-level information favor grouping of musical features, such as melodies and harmonies, over the details of their visual presentation. \linebreak MusicXML\cite{musicxml} is one such language. It is a markup language that describes score-level features in music in an application-independent way such that a composition can be passed between different score rendering programs.  Musical features are highly nested and groupings occur by visual features in the score. 
Unlike MIDI format, time is measured in MusicXML by measures and beats, which are relative to a particular tempo. 
\museci's implementation supports MusicXML as an input format to specify musical structures; the method for converting MusicXML's standard into our musical ECI data structures is described in section \ref{musicxml_conversion}.



\subsection{Algorithmic Score Analysis and Generation}

None of the languages discussed so far support operations for searching or manipulating the music.
MuSQL \cite{musql} is a structured query language for operating on musical \linebreak databases. It permits creation of, selection from, and alteration to musical structures. 
Its representations of musical features are generalized to address multiple kinds of pitch data, although time representations are limited to seconds---there is no method for describing more abstract metrical structures and containers such as measures and beats. 
\linebreak MuSQL uses standard SQL-style commands, which, although intuitive to programmers with the right training, are not necessarily easily derived from natural language statements. One of \museci's goals is to bridge that gap.

Music21 \cite{music21} is a Python library that has become popular for hypothesis testing on large musical copora.  It is able to read a variety of formats, including MIDI and MusicXML, and uses an object-oriented style that is very close to the structures used by MusicXML.  Music21 has been used in several computational musicology projects, including analysis of the large Bach chorales MusicXML corpus. However, Music21 is primarily structured based on music theoretic concepts, which, although potentially useful to analyzing questions about music cognitive models, is not always natural as a vehicle for mapping from natural language descriptions to operations on music.

Representations for score-level structures in programming languages intended for algorithmic composition typically revolve around the two classic computer science data structures of lists and trees. The Jython Music library \cite{jythonmusic}, a Python version of JMusic \cite{jmusic}, favors list representations for constructing melodies and chords using notes and rests. Melodies are lists of notes and rests with the assumption that one element's onset is determined by the previous element's end time. Chords are also represented as lists of notes having the same start time and duration. Python's support for nested lists allows for chords to be created within a larger sequential, melodic structure. While useful in pedagogical and algorithmic composition settings, this list-based approach makes representation of features such as arpeggiated chords tricky and its style of nesting musical concepts like parts and phrases is very rigid.

In the Euterpea library \cite{euterpea} for representing musical structures in Haskell, musical ideas are represented as binary trees, where leaves are notes and rests that store duration information but not onset time. The onset for a particular leaf is determined by its placement in the larger tree, whose intermediate nodes consist of two types of connectors: parallel and sequential. Sequentially composed subtrees are interpreted as following each other in time, and those composed in parallel are assumed to have the same start time (but not necessarily the same end time). Any pattern of notes that can be represented in the standard MIDI file format (which follows a list-based representation of time-stamped events) can also be represented as one of Euterpea's trees.\footnote{This is true for notes as they would appear on a score. Performance features, such as pitch bends and aftertouch, do not have a corresponding representation in Euterpea.} However, it is frequently the case that a given musical concept can have \emph{more than one} tree representation, making comparison of equality between tree structures and identification of shared features problematic. It may also not be cognitively natural to nest relationships in a strictly binary way---sometimes we may think of three or more items as having the same level of hierarchy, more like a list. Here, we use an n-ary tree representation (each node essentially being a list) to mitigate this issue in developing a framework to connect to NLP systems.

The structures and operations we use here are also very similar to some representations used in Kulitta \cite{icmc2015}, which is a Haskell-based framework for automated composition in multiple styles. This lends our implementation to being useful for automated composition tasks, while also allowing querying of data structures in a NLP-compatible way.

\subsection{Modeling Natural Language Semantics}

A key constraint for \museci is that it must accommodate the semantics of how music is described in natural language. To accomplish this, we take inspiration from computational languages designed for representing the natural language semantics of visual and physical environments, such as VoxML \cite{voxml}.

The VoxML modeling language encodes semantic knowledge of real-world objects represented as three-dimensional physical models along with events and attributes that are related and change through actions and interactions.  The framework encodes background knowledge about how interactions between objects affect their properties, allowing simulation to fill in information missing from language input. With this representation, natural language expressions describing scenes, events, and actions on objects can be specified in enough detail to be simulated and visualized.
We similarly seek a language in which musical concepts can be expressed in building-blocks that stand in close correspondence to natural language expressions.  The framework must capture enough detail to support expressing background music knowledge typically assumed in order to interpret natural language descriptions of music as well-defined musical structures that can be represented as scores and played---a kind of ``music simulation''.

\section{Musical Primitives}

Our lowest level of representations for Musical ECIs are called \emph{musical primitives}: concepts that represent fundamental musical units. Because we are focused on music at the level of a paper score, our primitives are based on features that appear on scores: notes and rests. 

\subsection{Symbolic Primitives}

\museci's 
atomic musical concepts are related to \linebreak pitch and duration, although these, by themselves, cannot be directly represented on a musical score. Currently we only consider Western tonal systems based on the chromatic scale and therefore represent pitch information using integers. Since some of the atomic concepts are best represented as numbers, we make use of two number types: $Int$ and $Float$.
We begin with the following two pitch-related atomic primitives, both of which are simply musical interpretations of numbers:
\begin{equation}
\begin{split}
& PitchClass(Int)\\
& Octave(Int)
\end{split}
\end{equation}

These then yield the following \emph{derived primitives}, which are built on atomic primitives.
\begin{equation}
\begin{split}
& Pitch(PitchClass, Octave)\\
& ScaleIndex(Int, Contexts)\\
\end{split}
\end{equation}

\noindent where $Contexts$ refers to a collection of environmental information and other labels used for resolving ambiguity. For $ScaleIndex$, the Context would need to include information about the current Scale. The \emph{pitch number} for a pitch can be computed by $PitchClass + 12(Octave + k)$, where $k$ has different standards in different areas of the literature.\footnote{Octave zero 
on a piano keyboard is not standardized and varies between musical programming languages. Sometimes $k=0$ such that (C,0) is pitch number 0, but it is also common to have $k=1$ and occasionally even $k=-1$ depending on the particular application.}

The following two atomic primitives are building-blocks for metrical information:
\begin{equation}
\begin{split}
& Beat(Float)\\
& Measure(Int)
\end{split}
\end{equation}

Derived primitives also exist for metrical information:
\begin{equation}
\begin{split}
& Onset(Measure,Beat)\\
& Duration(Beat)
\end{split}
\end{equation}

Pitch and metrical primitives are composed to produce the concepts of Note and a Rest, which are similar to the lowest levels of representation in Euterpea, JythonMusic and Music21. We define the set of all notes and rests to include the following primitives:
\begin{equation}
\begin{split}
& Note(Pitch, Duration, Onset, Contexts) \\
& Rest(Duration, Onset, Contexts)
\end{split}
\end{equation}

We denote the set of all such symbols as $P$. 
For brevity, in examples in this paper we will use $N$ to refer to the Python-style constructor for $Note$ and, in later sections, both pitch and metrical values will be shown only as tuples. An important difference between our approach and many other musical representations it that we do not require values to be declared for all properties of these concepts. The high degree of optional information in these constructs is needed to reflect the various ways in which we speak about music, which can involve many levels of abstraction and also may be ambiguous. For example, using ``\_'' to indicate a blank field, we may capture the concept of ``the C in measure 3'' as follows\footnote{In \museci, measure and beat numbers both index from zero. Therefore, ``measure 3'' produces a value of 2.}:
\begin{equation}
N(Pitch(0,\; \_), \; \_, \; Onset(2, \; \_), \; \_)
\end{equation}

This specifies a template that can be matched against a musical representation with more concrete information to identify the particular note being referred to. 
In MusECI, we represent ``\_'' using Python's \emph{None} value.
This variable degree of information is useful for specifying queries over musical structure and also for communicating musical ideas at a variety of levels of abstraction that may not have a uniform level of detail over the entire score. 

The Contexts field of Note and Rest indicates additional labels or other contextual information that may be useful for querying. Our implementation also supports volume as a note property, but it is omitted from the descriptions and examples in this paper.

\subsection{Connecting Primitives}

\museci uses Euterpea-inspired representations for connecting musical structures in sequence and in parallel. To minimize the issue of having multiple tree structures associated with Euterpea's binary trees, we use n-ary trees. This permits notes in a melody or chord that have the same level of conceptual hierarchy to appear at the same level within the representation. We also define normal forms for grouping structures from a score format such as MusicXML. The set of all possible sequential and parallel structures, which we will call $M$, is defined recursively.
\begin{equation}
\begin{split}
& Seq(\{M | P\}+, Contexts) \\
& Par(\{M | P\}+, Contexts)
\end{split}
\end{equation}

We will use the square bracket notation, $[...]$, to denote lists of items. Therefore, ``the three note melody in part B'' could be represented as:
\begin{equation}
Seq([N(\_), N(\_), N(\_)], \text{``Part B''})
\end{equation}

We will use the term \emph{Music value} to refer to any value that is in $M \cup P$. In other words, a Music value is some entity that could be drawn on a score, whether a single note, a rest, a melody, a chord, etc. Later we describe a collection of operations that can be applied to any of these values.

\subsection{MIDI and MusicXML Conversion}\label{musicxml_conversion}
\museci is able to parse both MIDI and MusicXML files into a normal form with the representations discussed so far. We use the following algorithm to convert MIDI and MusicXML information into our system's representations:
\begin{enumerate}
\item Greedily group Notes with the same onset and duration with $Par$.
\item Greedily group temporally adjacent structures (potentially including chords from step 1) with $Seq$.
\item Group any remaining temporally separated, but still sequential structures with $Seq$, using $Rests$ to fill temporal gaps. 
\item Group any leftover items under a global $Par$.
\end{enumerate}

Importantly, this normal form for reading MusicXML is not the only way to represent musical features. Other normal forms are possible, and we hope to improve our normal forms through learning in later work, such that nested structures within melodies and phrases are identified in genre-specific ways.

\section{Selection}

An important operation for composition by conversation is \emph{selection}. Much like the SELECT statement from SQL, which searches a database of tables for entries matching a query, we wish to scan over a musical data structure and return references to the correct portions of it.

We use \emph{pattern matching} as an integral part of our selection process, which involves checking to see whether concrete definitions match features present in an incomplete or abstract definition. 

\museci defines the $select$ operation as a function that takes two $Music$ values: one to use as a pattern to search for (a query) and the other to pattern match against. The ``\_'' values match any concrete value. For example, the statement
$select(N(\_, \_, ...), m)$
will select all notes from the $Music$ value,, $m$ and return a collection of references to the notes. Returning references in an object-oriented style allows for $m$, to be updated immediately after relevant parts of it are located: 

$query, operation = toQuery(parse)$

$operation(select(query, m))$

\noindent where $toQuery$ turns a parse of a natural language sentence into a query and an operation to perform on its result.

By leveraging Python's functional language features, \linebreak \museci also permits conditionals and classifiers as part of query statements. For example, finding notes above octave 3, can be expressed as $N((\_, >3), ...)$. Selection can also be done for more complex queries using $Seq$ and $Par$.

\section{Operations}

Support for music creation and generation requires that certain standard operators be defined to manipulate the musical data structures. The following are some examples of operations in \museci that we use in examples later on:

\begin{itemize}
\item $invert(Music)$ and $invertAt(Pitch, Music)$: generalized musical inversion (flipping upside down \linebreak along the pitch axis) at the first pitch and a specified pitch respectively.
\item $retrograde(Music)$: reverse a musical structure.  
\item $transpose(Int, Music)$: transpose a Music value by some number of half steps, either chromatically or according to scale degrees depending on the tonal context given. 
\end{itemize}

\section{Integration with TRIPS Parser}

\begin{figure}[ht]
\centering
\includegraphics[width=0.9\columnwidth]{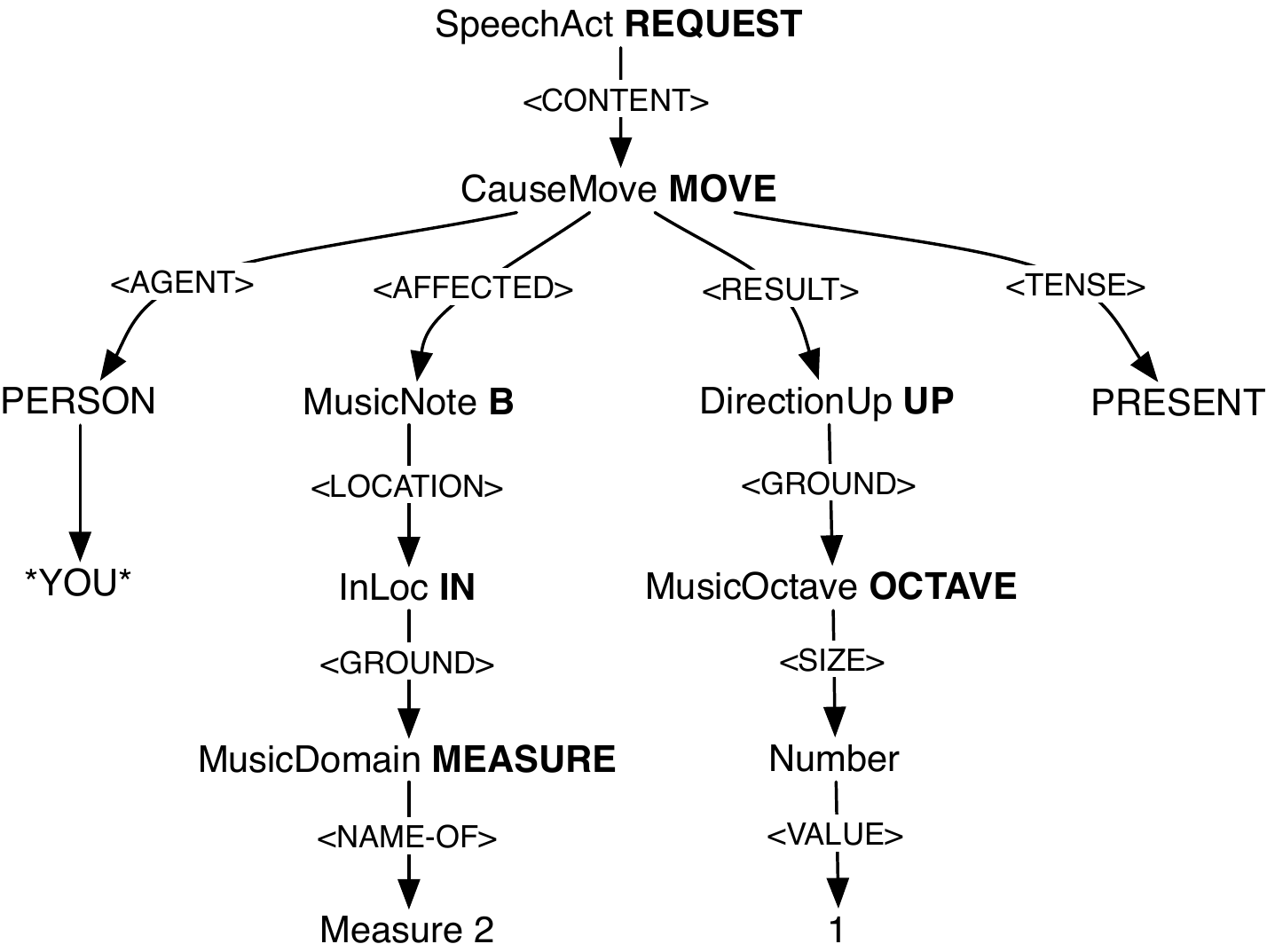}
\caption{TRIPS parser-like representation of the phrase ``Move the B in measure 2 up an octave.''}
\label{fig:trips-parse}
\end{figure}

In our work we use the TRIPS natural language parser.  TRIPS is a best-first bottom-up chart parser that integrates a grammar and lexicon
to encode both syntactic and semantic features.  These features are used to disambiguate and produce a \emph{logical form} representation that captures the semantic roles of terms in the utterance \cite{trips}.  Figure~\ref{fig:trips-parse} shows the logical form graph that TRIPS produces after parsing the phrase, ``Move the B in measure 2 up an octave.''  (Many details of the TRIPS logical form representation have been removed for presentation.)  Labels on arcs (surrounded by `$<...>$') represent semantic roles, and bold-face labels represent lexical terms preceded by their semantic type (e.g., {\bf B} is a MusicNote).

The logical form produced by TRIPS allows us to map from graphical patterns in the logical form to classes of operations on musical representations.  From the example in Figure~\ref{fig:trips-parse}, a {\sc Move} action in the direction {\sc Up} affecting a {\sc Note} corresponds to a transposition operation applied to our music representation.  The source location of the Note specifies the query for a note with a pitch class of B and an onset within measure 2.  Finally, moving up one octave corresponds to transposing up 12 half steps.  Together, this information specifies the following operation:
\begin{equation}
transpose(12, select(N((B,\_), \_, (1, \_)), m))
\end{equation}

\subsection{Resolving Ambiguities}

Natural language is extremely sensitive to conversational history. Sentences such as ``give that to him'' are impossible to semantically resolve without context for mapping pronouns to entities. In a musical setting, references such as ``the C next to it'' also suffer from this issue. 

Even with sufficient context to resolve references like pronouns, there may be insufficient detail to locate a precise portion of a musical concept. Consider the pitch classes in the opening refrain for ``Twinkle Twinkle Little Star:'' C, C, G, G, A, A, G. Requesting to ``move the G up a half step'' creates an ambiguity about \emph{which} G should be moved, since $select(N((G, \_), \_,...))$ will find three \linebreak matches. However, ``the G'' implies we only want to operate on one of them.

A system for addressing referential ambiguity requires two features: a working memory of recent references, and an \emph{assumer} algorithm or module that explores the working memory using features from a parse tree to \emph{resolve} ambiguities. That resolution step may be straightforward or it may involve requesting additional information from the user before proceeding. Disambiguation, therefore, has the following workflow, where $m$ is the target Music value.

$query, \; operation = toQuery(parse)$

$x = assumer.resolve(select(query, \; m))$

$operation(x)$

The creation of an appropriate assumer module is an area of ongoing work in conjunction with the TRIPS parser integration. Because of this, the examples presented in the next section are designed to avoid the two types of referential ambiguity described here. This means that all of the examples can be parsed directly to code statements of the form $operation(select(...))$ with no need for ambiguity resolution before applying the operation derived from the natural language sentence.


\subsection{Examples}

\begin{figure}[t]
\centering
\includegraphics[width=0.9\columnwidth]{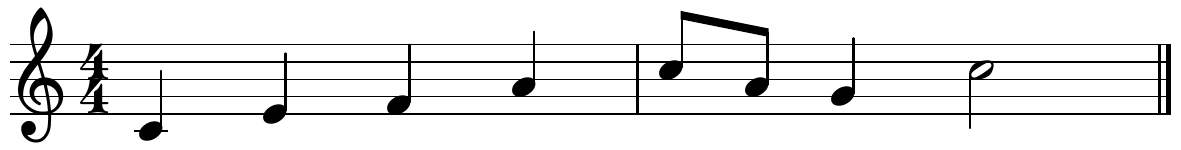}
\caption{Starting melody.}\label{melody1}
\end{figure}

\begin{figure}[t]
\centering
\includegraphics[width=0.9\columnwidth]{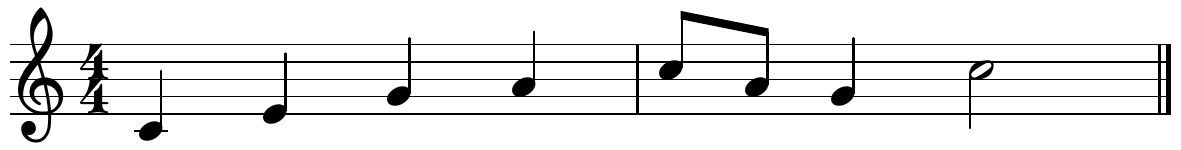}
\caption{The result of applying the change ``Move the F up a whole step'' to the melody in Figure \ref{melody1}.}\label{melody2}
\end{figure}

\begin{figure}[t!]
\centering
\includegraphics[width=0.9\columnwidth]{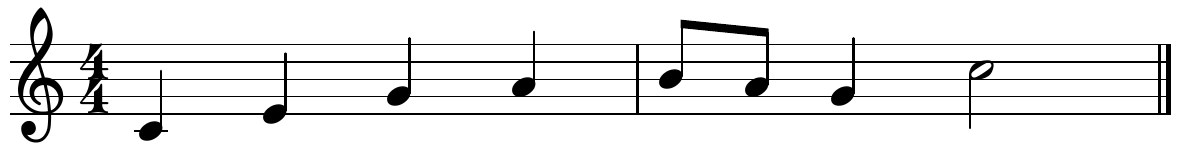}
\caption{The result of applying the change ``Move the C on the first beat of measure two down a half step.'' to the melody in Figure \ref{melody2}.}\label{melody3}
\end{figure}

\begin{figure}[t]
\centering
\includegraphics[width=0.9\columnwidth]{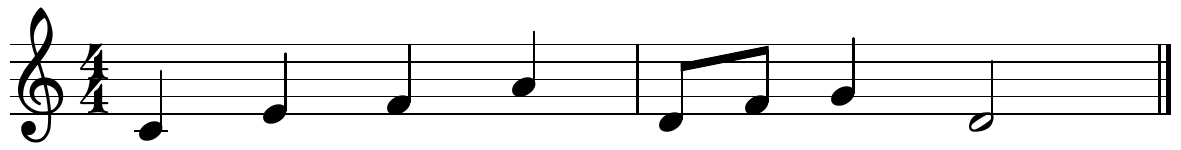}
\caption{The result of applying the change ``invert the notes in measure two around G4'' to the melody in Figure \ref{melody2}.}\label{melody4}
\end{figure}

\begin{figure}[h!]
\centering
\includegraphics[width=0.9\columnwidth]{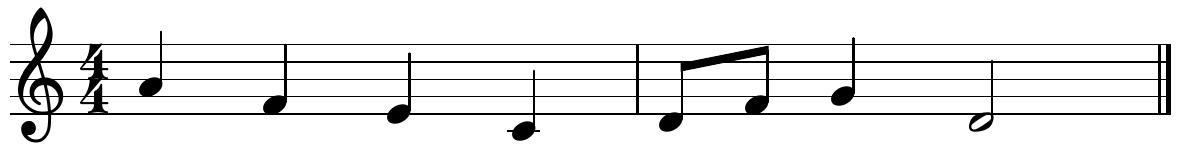}
\caption{The result of applying the change ``reverse the notes in measure one'' to the melody in Figure \ref{melody4}.}\label{melody5}
\end{figure}

\begin{figure*}[th!]
\centering
\includegraphics[width=15cm]{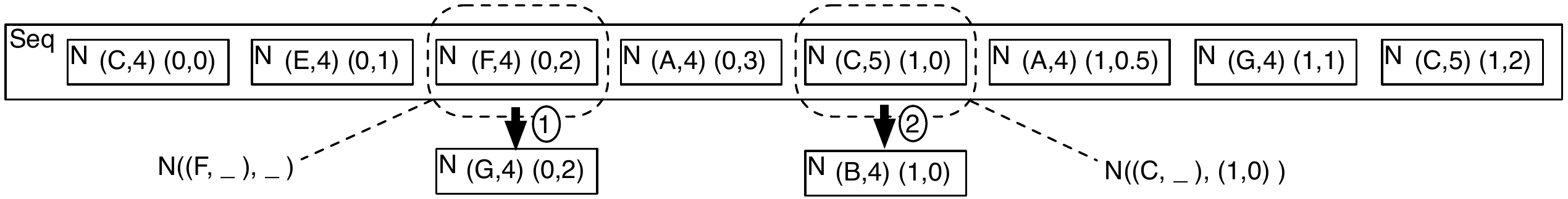}
\caption{Data structure representation of the melody in Figure \ref{melody1} and the two alterations produced by the operations from conversation 1 that yield the melodies in Figures \ref{melody2} and \ref{melody3} respectively. Duration information is not referenced directly by any of the selection operations and is therefore omitted from notes in this diagram; \textsf{N} is shorthand for a Note object and constructor. }\label{datastructure}
\end{figure*}

\begin{figure*}[th!]
\centering
\includegraphics[width=15cm]{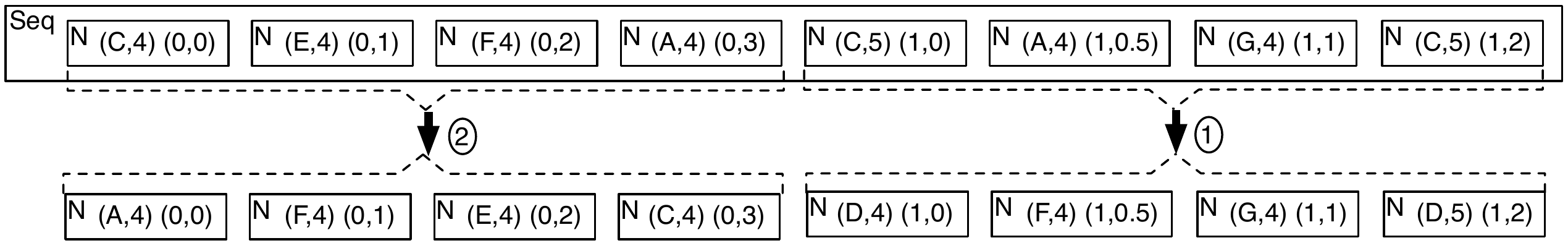}
\caption{Data structure representation of the melody in Figure \ref{melody1} and the two alterations produced by the operations from conversation 2 that yield the melodies in Figures \ref{melody2} and \ref{melody3} respectively. Duration information is omitted for the same reasons as in Figure \ref{datastructure}. }\label{datastructure2}
\end{figure*}

Consider the two-measure melody shown in Figure \ref{melody1}, and suppose we wish to transform it into the melody shown in Figure \ref{melody3} through natural language commands. A precise way to describe this 
would be ``move the F up a whole step'' followed by ``move the C on the first beat of measure two down a half step.'' This will create the following interactions between the computer (C) and user (U). \\

\textbf{Conversation 1}

C: $m$ = the melody in Figure \ref{melody1}.

U: ``Move the F up a whole step.''

C: $transpose(2, select(N((F, \_), \_, \_, ...), m))$

U: ``Move the C on the first beat of measure two down a

half step.'' 

C: $transpose(-1, select(N((C, \_), \_, (1,0), ...), m))$ \\

Applied sequentially, these commands produce the \linebreak melodies shown in Figures \ref{melody2} and \ref{melody3} respectively. The data structure for the starting melody is shown in Figure \ref{datastructure}, along with the selection statements created by each of the commands and the resulting modification to the data structure. 

Statements may also operate over collections of notes. The following conversation illustrates this with inversion and reversal of sections of the music (retrograde). Figures \ref{melody4} and \ref{melody5} show the result of each command in this conversation, and the data structure transformation is shown in Figure \ref{datastructure2}. \\ 

\textbf{Conversation 2}

C: $m$ = the melody in Figure \ref{melody1}.

U: ``Invert the notes in measure two around G4.''

C:
$invertAt((G,4), select(N(\_, \_, (1,\_), ...), m))$

U: ``Reverse the notes in measure one.''

C: $retro(select(N(\_, \_, (0,\_), ...), m))$

\section{Conclusion and Future Work}

We have proposed the \museci framework for modeling music in a way that allows querying based on parsing of natural language statements. Our Python implementation of the musical data structures supports normal algorithmic composition in a text editor in addition to the composition by conversation use case in a way that is easily integrated with parsing systems like TRIPS. 

Although our data structures are robust for representing complex musical structures, the overall system is currently a prototype limited to a fairly narrow collection of musical operations. This collection of operations needs to be expanded to feature more tree manipulation algorithms, such as different strategies for removing and re-arranging notes within a larger structure. Operations such as removal of individual notes can have several potential interpretations, such as replacement with a rest of the same duration and removal followed by time-shifting other elements of the data structure to close the gap. Both operations are valid, but which is most appropriate depends on the larger context in which it is used. We are currently working on a more diverse array of operations such as this to support a wider range of common score manipulations.

Expansion of this framework for NLP-based music composition and manipulation will require adaptation of NLP toolkits like the TRIPS parser. Standard parsers often do not have a suitable dictionary of musically-relevant definitions to draw on when assigning semantics to nouns and verbs. While terms like ``transpose'' and ``reverse'' can be interpreted correctly due to their usage in non-musical settings, correct interpretation of other terms is trickier. Without incorporation of music-specific terminology, the TRIPS parser identifies``C'' and ``F'' as temperature scales rather than as pitch classes. Just as humans learning about music must have their vocabulary expanded, we are developing a music-specific ontology that can be incorporated in the TRIPS parser to support a more diverse range of musical concepts and operations to achieve a correct parse tree for sentences in a composition by conversation setting.

The addition of new representations for contexts is important for the creation of an assumer module. This requires a way to infer the referents of ambiguous words like ``it,'' ``this,'' and so on as well as keeping track of what spaces or metrics are currently in use. 

Our prototype framework is the beginning of an integrated approach for handling natural language and musical concepts. Currently, \museci and its integration into NLP systems is still a work in progress. However, we hope to eventually achieve real-time interactive programs capable of interacting with real musicians in a musical setting, similar to how AI on computers and mobile devices is currently able to respond to basic spoken commands in other domains.  We also aim for this communication to ultimately become bi-directional and incorporate other aspects of musical artificial intelligence, perhaps even allowing the machine to critique its human user's musical ideas or offer its own suggestions to help complete a musical project.


\bibliography{musicbib3}

\end{document}